\documentclass{nature}
\usepackage{amsmath}
\usepackage{graphicx}
\usepackage{hyperref}
\usepackage{siunitx}

\title{35 megawatt multicycle THz pulses from a homemade periodically poled macrocrystal}

\author{Fran\c{c}ois Lemery$^{1}$, Thomas Vinatier$^1$, Frank Mayet$^1$, Ralph A{\ss}mann$^1$, Elsa Baynard$^{2}$, Julien Demailly$^{3}$, Ulrich Dorda$^{1}$, Bruno Lucas$^{3}$, Alok-Kumar Pandey$^{3}$ \& Moana Pittman$^{2}$}

\makeatletter
\let\saved@includegraphics\includegraphics
\AtBeginDocument{\let\includegraphics\saved@includegraphics}
\renewenvironment*{figure}{\@float{figure}}{\end@float}
\makeatother
\begin{document}

\maketitle

\begin{affiliations}
 \item DESY, Notkestrasse 85, 22607 Hamburg, Germany
 \item LASERIX - LUMAT (FR2764), University Paris-Saclay, France
 \item LPGP (UMR 8578), University Paris-Saclay, France
\end{affiliations}

\begin{abstract}
High-power multicycle THz radiation is highly sought after with applications in medicine\cite{medical}, imaging\cite{tomography}, spectroscopy\cite{spectroscopy}, characterization and manipulation of condensed matter~\cite{THzStrong}, and could support the development of next-generation compact laser-based accelerators with applications in electron microscopy, ultrafast X-ray sources\cite{lemeryTaper,vinatier} and sub-femtosecond longitudinal diagnostics\cite{lemeryStreak}.  Multicycle THz-radiation can be generated by shooting an appropriate laser through a periodically poled nonlinear crystal, e.g. lithium niobate (PPLN)\cite{leethzppln,leethztemp, leeTunable, leeShaping,vodo,sergio,jolly2019}.  Unfortunately, the manufacturing processes of PPLNs require substantially strong electric fields $\mathcal{O}(10~kV/mm)$ across the crystal width to locally reverse the polarization domains; this limits the crystal apertures to well below 1~cm.  Damage threshold limitations of lithium niobate thereby limits the laser power which can be shone onto the crystal, which inherently limits the production of high-power THz pulses.  Here we show that in the THz regime, a PPLN crystal can be mechanically constructed in-air by stacking lithium niobate wafers together with 180$^{\circ}$ rotations to each other.  The relatively long (mm) wavelengths of the generated THz radiation compared to the small gaps ($\sim$10~$\mu$m) between wafers supports a near-ideal THz transmission between wafers.  We demonstrate the concept using a Joule-class laser system with $\sim$50~mm diameter wafers and measure up to 1.3~mJ of THz radiation corresponding to a peak power of $\sim$35~MW, a 50 times increase in THz power compared to previous demonstrations\cite{jolly2019}.  Our results indicate that high-power THz radiation can be produced with existing and future high-power lasers in a scalable way, setting a course toward multi-gigawatt THz pulses.  Moreover the simplicity of the scheme provides a simple way to synthesize waveforms\cite{leeShaping} for a variety of applications.  We expect our results to have a broad range of appeal, including non-linear optics, high-power pump-probe spectroscopy\cite{xfelTHz} and to illuminate a path toward laser-based table-top high brightness particle accelerators and diagnostics.
\end{abstract}
THz radiation is an attractive candidate to power next-generation laser-based particle accelerators, where the millimeter-scale wavelengths and large accelerating fields (100+ MV/m) could support the generation of femptosecond electron bunches on compact footprints.
Two prevailing paths toward powering THz-based accelerators have emerged. Cavity-based approaches seek to fill accelerating cavities with narrowband pulses\cite{nanniProto,nanniTHz,moeinGun}; over the fill time, the accelerating field strength increases until a properly timed electron bunch is accelerated.  This approach requires appropriate bandwidths ($<$1\%) to match the quality factor of the cavity.

Alternatively, travelling wave approaches in e.g. dielectric-lined waveguides do not rely on filling times, rather they use the fields directly\cite{nanniNature, emmaZero,lemeryTaper,lemeryStreak,vinatier}.  Here the duration of the THz pulse and group velocity in the waveguide limit the interaction length and energy gain; consequently, high THz peak-powers are advantageous to produce large accelerating gradients. 

Laser-based narrowband THz generation has been thoroughly investigated with periodically poled crystals\cite{QPM,leethzppln,leethztemp, leeTunable, leeShaping,vodo,sergio,jolly2019, raviCascade,raviSequence}, also in e.g. parametric oscillators\cite{thzOPO}.  
For high-power applications, the limited aperture sizes of periodically poled crystals ($<$1~cm) has motivated the development of wafer-bonded quasi phase matching crystals\cite{waferbonding,leewafer}. However, these fabrication techniques are challenging for large diameter wafers and require delicate in-vacuum procedures, limiting their widespread use.
\begin{figure}
\centering
\includegraphics{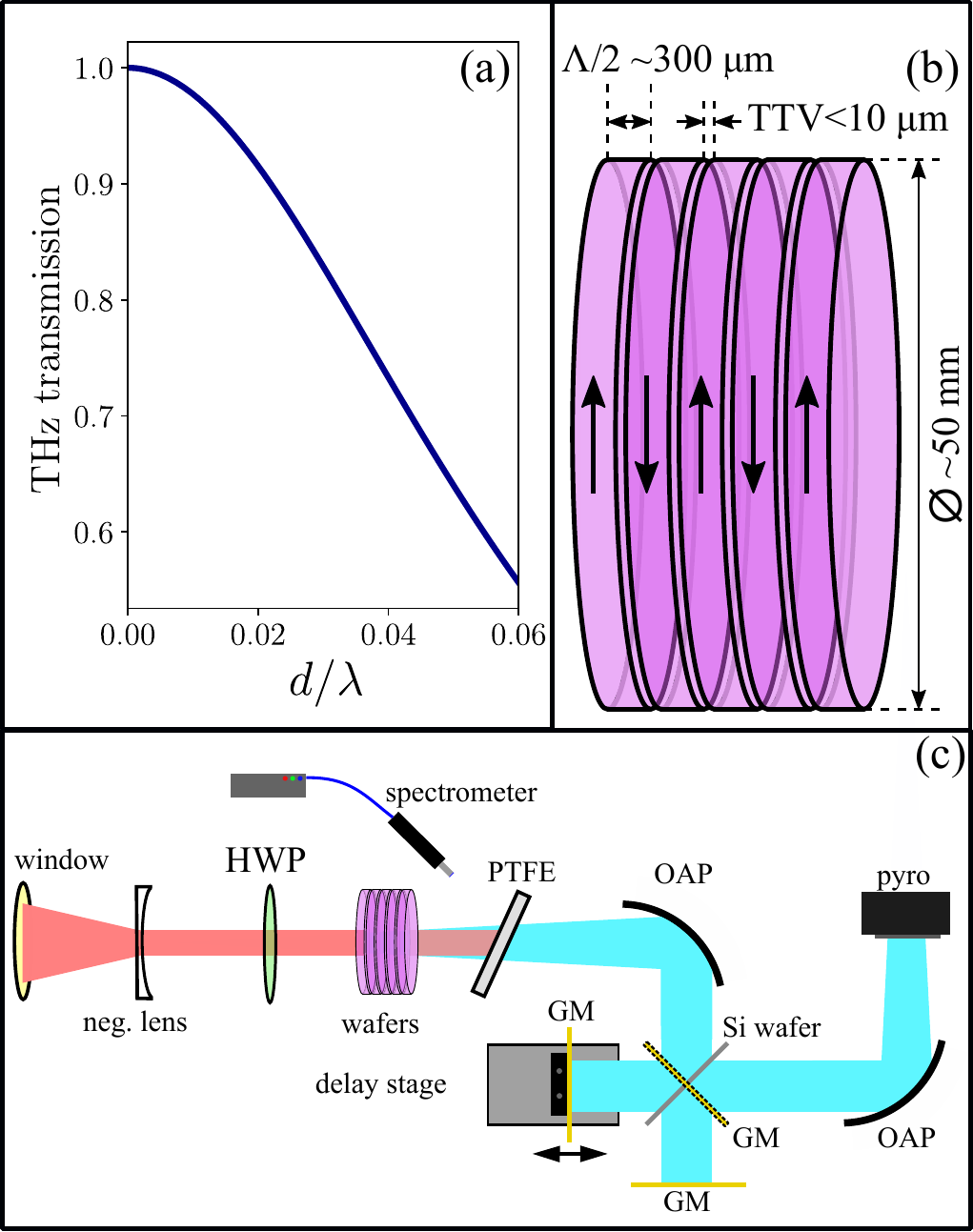}
\caption{The THz transmission between wafers as a function of the ratio between the wafer separation and THz wavelength ($d/\lambda$) is illustrated in (a) corresponding to Eq.~\ref{eq:tranny}. A schematic of the homemade PPLN macrocrystal is shown in (b); the wafers are successively rotated by $\pi$ for quasi phase matching.  (c): Diagram of the experimental setup.  Here HWP=half-wave plate, GM=gold mirror, OAP=off-axis parabolic mirror, pyro=pyroelectric detector.  The setup could be easily modified for energy measurements (using dashed-GM, no Si wafer), and for an autocorrelation measurement (no dashed-GM, with Si wafer beam splitter.)  Here the red beam represents the ti:saph pump laser, and the blue represents the generated THz beam.}
\label{fig:stack-setup}
\end{figure}

Alternatively, the relatively long (mm) wavelengths of THz radiation supports a simple fabrication of a wafer stack in simple laboratory conditions.  Just as a large wheel hardly feels a small bump on the road, here the generated THz wavelengths are much longer than the separation between wafers, supporting very small reflection coefficients at the wafer interfaces. Fig.~\ref{fig:stack-setup}(a) shows the transmission between two wafers as a function of the separation distance per wavelength $d/\lambda$, and Fig.~\ref{fig:stack-setup}(b) depicts the assembled wafer stack; see Methods for a detailed mathematical description.

The domain period $\Lambda$ (which is two times a wafer thickness), along with the refractive indices of lithium niobate for THz ($n_{THz}\sim5.05$) and IR ($n_{IR}\sim2.2$) determine the central frequency of the THz radiation via, 
\begin{equation}
    f_{THz}=\frac{c}{\Lambda(n_{THz}-n_{IR})}.
\label{eq:thzfreq}
\end{equation}

The experiment was carried out at the LASERIX facility of Paris-Saclay Univeristy at Orsay~\cite{laserix2}, the laser parameters are presented in Table \ref{tab:laserix} in Methods. The LASERIX laser driver is a chirped pulse amplification~\cite{CPA} based Ti:Sa system which provides different femtosecond beamlines from the mJ-level to $\sim$1.2~J after compression. All beams operate at 10 Hz and the main beam can be compressed down to 50~fs with the in-vacuum movable grating compressor.

While conventional lasers generally produce transverse Gaussian mode profiles, high-energy multi-amplification systems tend to super-Gaussian (SG) profiles due to saturation effects with flat-top pump-laser profiles.  While SG beams are more challenging to transport, the increased uniformity of the transverse distribution of the intensity is useful for THz generation, since the IR-to-THz conversion efficiency is proportional to the intensity squared, leading to a higher overall conversion efficiency while remaining below intensity-limited damage thresholds of the wafers.  

Moreover, the propagation dynamics of the generated THz beam depend strongly on the beam size of the pump laser beam, the half-angle of the THz divergence goes as $\theta \approx \lambda_{THz}/\pi n_{THz} w_0$, where $w_0$ is the beam size.  In this respect, larger laser beam sizes in the crystal are significantly advantageous, allowing for efficient waveform generation along the wafer stack.

Fig.~\ref{fig:stack-setup}(c) displays a diagram of the experimental setup.  The main beam was rerouted into air by placing a converging lens of $f=3~m$ in the vacuum chamber to reduce the beam size to pass through the 2 inch MgF$_2$ window.  MgF$_2$ is chosen for its low dispersion and low non-linear index n2 which are appropriate for femtosecond applications.  A diverging lens was appropriately placed in air to provide a magnification $M=0.5$.  A zero order half-wave plate was used to adjust the polarization of the IR beam with respect to the extraordinary axis of the wafers.  The wafer stack was fabricated by hand in air, see Methods. After propagation through the wafer stack, the IR beam was dumped onto a a white diffuser supported on a rectangular PTFE mount.  The spectrum of the IR was captured with a spectrometer, the fiber coupler was pointed at the diffuser.  The THz radiation propagated through the PTFE and went to our THz-diagnostics setup.

Two THz diagnostics were used (see Fig.~\ref{fig:stack-setup}(c)). The first was dedicated to measuring the THz energy and consisted of two off-axis parabolic mirrors (OAP) placed to provide 4F imaging into a THz pyroelectric detector.  A second diagnostic was constructed to measure an autocorrelation of the THz signal.  Details of this diagnostics can be found in Methods.  

\begin{figure}[b!]
\centering
\fbox{\includegraphics[width=\linewidth]{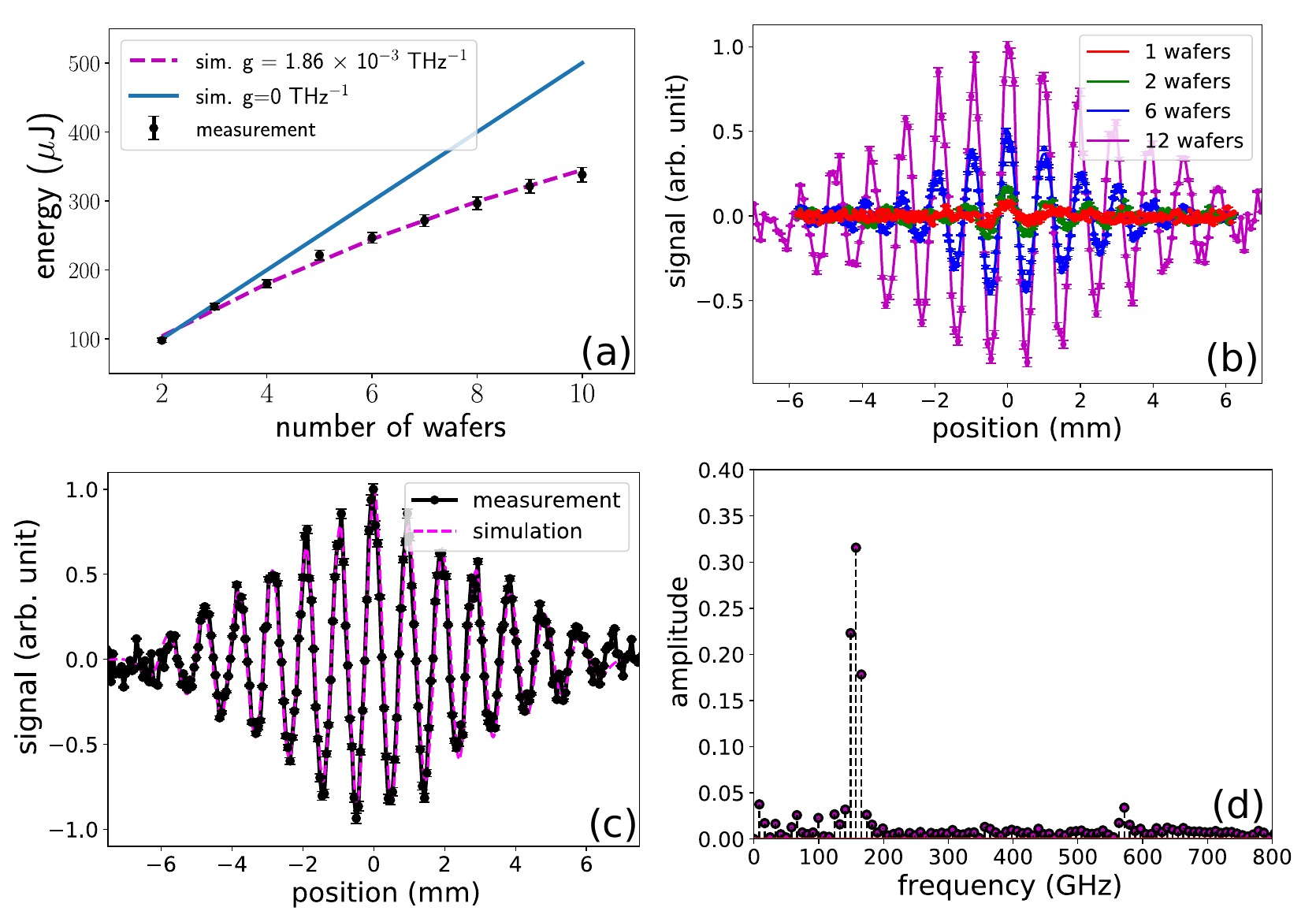}}
\caption{In (a) the THz pulse energy scaling as a function of the number of wafers is illustrated.  The absorption $g$=\num{1.86e-03} THz$^{-1}$ is fitted with Eq~\ref{eq:thzEqn} (see Methods).  In (b), autocorrelation traces for 1, 2, 6 and 12 wafers in the stack are presented.  (c) shows the agreement between Eq.~\ref{eq:thzEqn} and an autocorrelation trace for 12 wafers (the absorption coefficient fitted from (a) is used).  Finally, (d) illustrates the THz pulse frequency spectrum (Fourier transform of (c)), with a central frequency $f_0\sim$160~GHz.}
\label{fig:fan4}
\end{figure}

The output energy and evolution of the THz waveforms for stacks with different numbers of wafers was investigated. 
Initial wafer tests indicated some differences between individual wafers.  The wafers were accordingly sorted (see Methods).  Subsequently, we investigated the performance of the wafers with a pump laser energy of $\mathcal{E}$=267~mJ, rms pulse duration of $\tau\sim$700~fs and a FWHM beam diameter of 18~mm.  The resulting energy measurements are illustrated in Fig.~\ref{fig:fan4}(a) with black markers. The absorption coefficient in the stack $g$, was calculated from the fit of the data (magenta trace), and Eq.~\ref{eq:thzEqn} (see Methods), to $g$=\num{1.86e-03} THz$^{-1}$.  Autocorrelation signals for 1, 2, 6, and 12 wafers are shown in Fig.~\ref{fig:fan4}(b). Fig.~\ref{fig:fan4}(c) illustrates the comparison between measurement (black) and simulation (magenta) using the calculated absorption factor $g$ and Eq.~\ref{eq:thzEqn} (see Methods). Finally, the power spectrum is shown in Fig.~\ref{fig:fan4}(d) for a stack of 12 wafers; the central frequency occurs at $f_{THz}\sim$160~GHz, in good agreement with Eq.~\ref{eq:thzfreq}.

\begin{figure*}[t]
\centering
\fbox{\includegraphics[width=\linewidth]{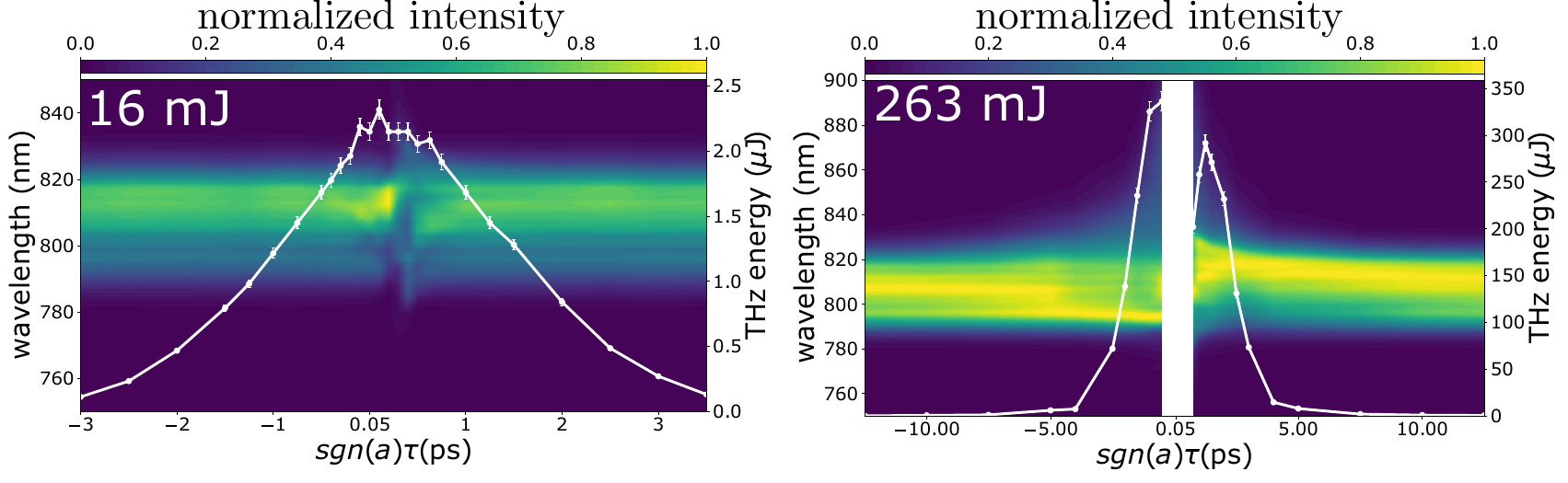}}
\caption{Spectra of the IR as a function of the effective pulse length after 12 wafers for 16~mJ and 263~mJ of pump energy. Here the IR grating compressor was moved through a large range of compression yielding positive and negative chirps ($a$), see Methods. The white trace (right vertical axis) corresponds to the measured THz energy.  For the case of 263~mJ, we avoided measurements below $\tau$=$\sim$700~fs due to filamentation in the negative lens.}
\label{fig:spectra}
\end{figure*}

We explored the IR spectrum after 12 wafers for various pulse durations (chirps) with 16 and 263~mJ of pump energy with a FWHM beam diameter of 18~mm (see Fig.~\ref{fig:spectra}).  For the 263~mJ case we were limited by filamentation in the negative lens for pulse durations below $\tau \sim$700~fs. For the case of 16~mJ, we were able to explore the maximum pulse compression range of $\tau \sim$50~fs.  At large pump intensities, a significant redshift was observed along with some smaller amounts of blueshift for negative chirps.  The simple intensity-based model from Eq.~\ref{eq:thzEqn} (see Methods) does not include cascading effects. More elaborate theoretical models from K. Ravi~\cite{raviSequence,raviCascade} or~\cite{vodo} study these intricate behaviors in detail.  We note that such behaviors can also come from other non-linear effects in the wafers such as self-phase modulation.

\begin{figure}[h]
\centering
\fbox{\includegraphics[width=\linewidth]{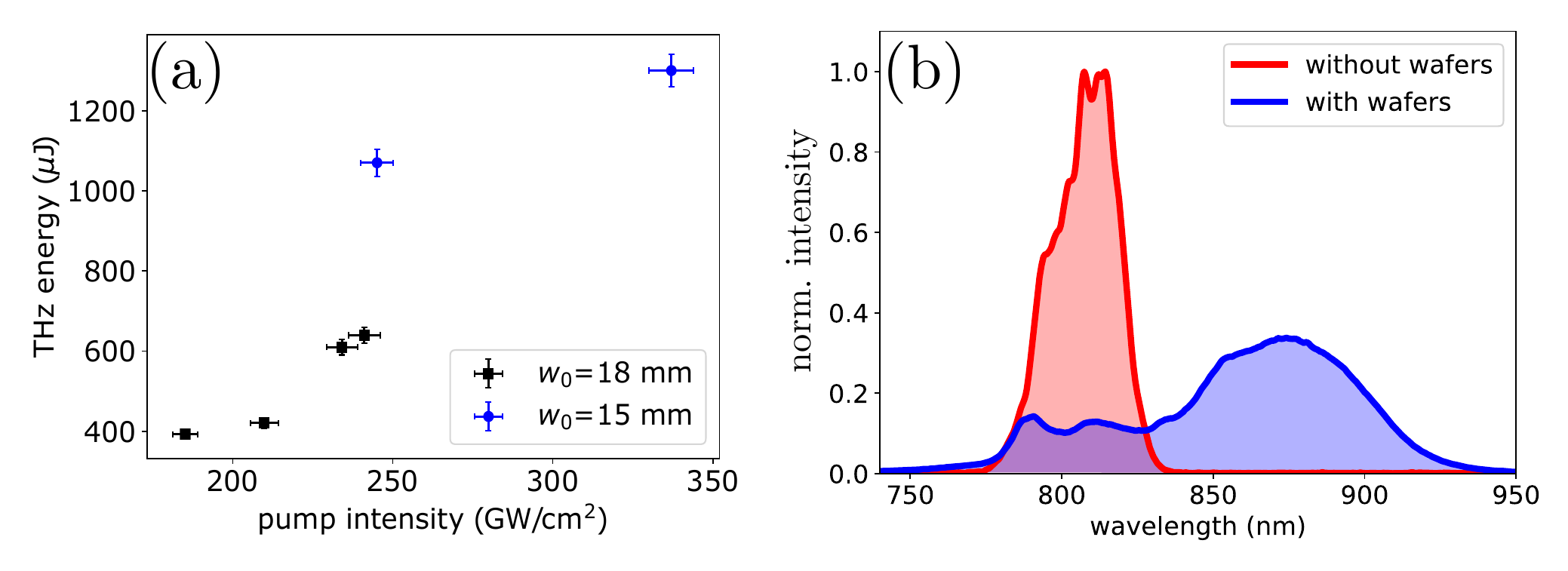}}
\caption{(a): THz energy as a function of pump intensity for various pump energies and pulse durations. (b): Spectra of the IR without (red) and with the wafers (blue) is shown; the center-of-mass shifts from 808~nm to 858~nm.}
\label{fig:highE}
\end{figure}

The limitations of the wafer stack with increasing IR pulse energies were also investigated.  We proceeded with a FWHM IR beam diameter of 18~mm and rms pulse duration of $\tau$=1.25~ps, the THz energy as a function of pump intensity is illustrated in Fig.~\ref{fig:highE}(a).  Above 1.1~J, filamentation occurred in the negative lens and we therefore increased the pulse duration and took a further two data points with $\tau$=1.4~ps and $\tau$=1.6~ps.

With no signs of damage from the LN wafers, we subsequently removed the negative lens from our experimental setup to reduce the FWHM IR beam diameter to 15~mm over the long, 3~m focus.  The pump energy was limited to $\mathcal{E}$=910~mJ and we performed two measurements with different pulse durations, $\tau$=1.65 and $\tau$=1.2~ps; the corresponding measured THz energies were, respectively, $\mathcal{E}_{THz}$=1.07 and 1.3~mJ, the latter corresponding to a peak power of $\sim$35~MW.  The data points are shown in Fig.~\ref{fig:highE}(a) (blue stars), the jump in the generated THz energy for similar intensities with a smaller beam size is clearly attributed to the limited collection efficiency in our 4F diagnostics.
The associated redshift for the 1.3~mJ case is shown in Fig.~\ref{fig:highE}(b), a center-of-mass shift of $\sim$50~nm is exhibited. From energy conservation, an upper limit on the IR-to-THz conversion efficiency can be calculated from the spectra shift via $\eta_{IR}=\frac{\bar{\nu_i}-\bar{\nu_o}}{\bar{\nu_i}}$, where $\bar{\nu_i}$ and $\bar{\nu_o}$ are the frequency center-of-mass of the IR spectrum before and after the stack respectively.  For Fig.\ref{fig:highE}(b), the redshift corresponds to a maximum efficiency of $\eta$=5.8\%.

There is a drastic difference between the measured conversion efficiency, $\eta$=0.14\% and the spectrum-calculated maximum efficiency $\eta_{IR}$=5.8\%.  Several factors can account for this discrepancy.  First, the 4F imaging in our setup was not ideal for the limited aperture size of the THz pyroelectric detector ($9$~mm). With the anticipated super-Gaussian transverse profile of the THz, we estimate a factor of $\sim$3 was not measured by the detector (see Fig.~\ref{fig:highE}(a)).  Second, the large refractive index for the THz in LN ($n_{THz}=5.05$) generates a large reflection at the output surface into air, $\mathcal{R}$=0.45.  THz absorption in the wafers (as visible in Fig~\ref{fig:fan4}(a)) also reduces the measured energy by $\sim$31\%.  Finally, as discussed in ~\cite{raviCascade}, the generated THz is also inadvertently phase matched with the IR, leading to some THz energy being absorbed and blueshifting the IR spectrum.

Another experiment is being planned to incorporate significant improvements to the experimental setup, including a better THz transportation, cryogenic cooling of the wafers to minimize absorption losses~\cite{leethztemp},  output coupling using e.g. fused silica wafers to decrease the reflection at the end of the wafer stack, and an investigation of significantly longer wafer-stacks.  The future experiment will also be used in preparation for an experiment aiming at injecting the THz produced with LASERIX into a circular dielectric-lined waveguide and post-accelerate an electron bunch produced by the photoinjector PHIL located at LAL~\cite{PHIL}.  Finally we note that the super-Gaussian profiles are suspected to be beneficial to improve mode coupling into dielectric-lined waveguides, where the transverse field profiles are uniform for phase velocities $v_p=c$ (see~\cite{lemeryTaper,vinatier}).


\begin{methods}
\subsection{THz generation in PPLN}
The generated field in a PPLN can be described well~\cite{leethzppln} via the partial differential equation,    $\big(\frac{\partial^2}{\partial z^2}+ n^2(\omega)\frac{\omega^2}{c^2}\big) E(z,\omega)=-\mu_0 \omega^2 P(z,\omega)$, where $z$ is the longitudinal propagation coordinate, $\omega$ is the angular frequency, $c$ the speed of light, $\mu_0$ the magnetic vacuum  permeability, $E(z,\omega)$ the electric field, and $P(z,\omega)$ is the nonlinear polarization density of the medium.  The electric field can be calculated as
\begin{equation}
E_\text{THz}(z=L,t)=\frac{1}{L}\sum_{i=1}^N(-1)^{i-1}\int_{l(i-1)}^{l(i)}E_\text{THz}^\text{local}(z', t)dz',
\label{eq:thzEqn}
\end{equation}
where $E_{THz}^\text{local}=\frac{\pm A_0\tau}{4\eta(z')^{3/2}}(\frac{(t-t_d(z'))^2}{2\eta(z')}-1) \exp(-\frac{(t-t_d(z'))^2}{2\eta(z')})$,  $t_d(z')=(n_\text{IR}z'+n_\text{THz}(L-z'))/c$, $\eta(z')=\frac{\tau^2}{4}+\frac{g}{c}(L-z')$, $\tau$ is the IR rms pulse duration, $L$ is the stack length, $N$ is the number of wafers, $A_0$ is the field amplitude and $g$ is the decay time associated to losses.

\subsection{THz reflections at wafer boundaries}
For normal incidence, the THz transmission coefficient between each wafer can be calculated from \cite{jacksonEM}, 
\begin{equation}
    T=\frac{4n_\text{THz}^2}{4n_\text{THz}^2+(1-n_\text{THz}^2)^2\sin(\frac{2\pi d}{\lambda_\text{THz}})^2},
    \label{eq:tranny}
\end{equation}
where the THz wavelength, $\lambda_\text{THz}=\frac{c}{f_\text{THz}}$, $f_THz$ is the corresponding THz frequency and $c$ is the speed of light in vacuum.  The transmission as a function of the ratio between the wafer spacing and generated wavelength, $d/\lambda$, is illustrated in Fig.~\ref{fig:stack-setup}(a). As an example, for $d/\lambda=0.001$, $T\sim 0.999$.

\subsection{Wafer specifications}
The wafers were provided by Precision Micro Optics, see Table~\ref{tab:wafers} of Methods for specifications.    When stacking wafers together, good surface contacts were evident from adhesion forces between the wafers i.e. the wafers could not be separated by pulling, rather only by sliding the wafers along the surface faces. The wafers had alignment markers to distinguish the polarization axes and were accordingly stacked and placed into an optical mount. 
The total thickness variation (TTV)$<10~\mu$m of the wafers indicated an expected transmission coefficient from wafer-to-wafer of $\sim$99.4\% (Eq.~\ref{eq:tranny}).
To overcome the reflection of the pump-laser operating at significantly smaller wavelengths ($\sim$800~nm), each wafer face had an anti-reflective coating which provided a transmission above 99.75\% per face, see Fig.~\ref{fig:stack-setup}(b) for a schematic of the wafer stack.

\subsection{Wafer sorting}
In early tests, we observed different responses from individual wafers and sought to identify the reason for this behavior.  We sorted the wafers by measuring the polarization rotation of the pump-laser by using an energy detector placed after a polarizing cube.  X-cut wafers should not have any birefringence for polarizations aligned in the z-plane, according to our purchase order; however, this was not the case.  We distinguished approximately 3 distinct families of wafers from our order; there were a small amount of outliers which we discarded.  Each wafer was tagged with a non-permanent marker.  The remainder of all measurements were conducted with the `purple' family, which exhibited the expected response.
We also sorted the wafers in terms of thickness to determine if there was any contribution to our observations using a micrometer-caliper.  We confirmed that the thickness variation was not responsible for the observed differences by mixing two families with the same thickness.  Moreover, the thickness variation was within specifications.  We suspect the wafers we received from the manufacturer were not properly cut.


\subsection{Wafer stacking}
To minimize THz reflections at wafer-wafer boundaries, the wafer separation distance should also be minimized according to Eq.~\ref{eq:tranny}.  Each wafer was thoroughly cleaned with acetone and optical-wipes to remove peculiarities which came from the manufacturer.  We explored various ways to minimize the contact between wafers.  Our most successful approach was to slide the wafers onto each other in parallel.  This avoided the formation of air-pockets by contacting two wafers directly.  Good wafer contacts were evident from the adhesion forces between the wafers.  We tried using various optical mounts to support the wafers.  We used a typical FMP2 mount from Thorlabs, but the rotation clamp led to the rotation of some wafers. This was solved by applying some tape to the sides of the wafer stack.  We also used a simple c-clamp to hold the wafers together and we did not notice any change in the signal strength.

\subsection{Pump and THz separation}
The high-power laser pump and the THz need to be separated for THz diagnostics and future applications.  We tested various approaches but found that a white optical cloth diffuser attached to a 1~cm thick PTFE block worked very well.  The diffuser scattered the pump beam without affecting the THz; without the diffuser, the PTFE began taking damage quickly.

\subsection{Autocorrelation}
The autocorrelator was assembled with two gold off-axis parabolic mirrors (OAPs), two planar gold mirrors, a silicon wafer and a THz detector, see Fig.~\ref{fig:stack-setup}.  The input THz beam is directed onto a silicon wafer, which serves as a beam splitter. The reflected portion is subsequently reflected off of a gold mirror mounted on a linear translation stage, the transmitted portion of the THz beam is also reflected on another gold mirror.  Both the transmitted and reflected parts are recombined on the beam splitter and transported to the pyroelectric detector.  The interference between both paths leads to the autocorrelation signal.
The autocorrelation signal strength is limited by the transmission and reflection coefficients on the silicon wafer.  By exchanging the silicon beam splitter with a gold mirror, we were able to measure the THz energy directly, see Fig.~\ref{fig:stack-setup}.

\subsection{THz detectors}
All three used THz detectors were from GentecEO.  A QE9SP-B-MT-L-BNC-D0 was used for low-energy measurements and is capable of measuring approximately 100~nJ of THz radiation.  A second detector, QE8SP-B-MT-D0 is much less sensitive but capable of measuring into the mJ range.  This was used for the majority of the measurements.  Finally, a third calibrated THz detector was used, THZ5I-MT-BNC to cross-calibrate the other two detectors for our energy measurements.

\subsection{IR pulse measurements and calibration}
Laser pulse durations  $\tau <$ 100 fs were measured using self-referenced spectral interferometry (Wizzler, Fastlite), whereas for 100$<\tau<$1000 fs, a single-shot autocorrelator (TiPA, Light Conversion) was used. Pulse durations longer than 1 ps were estimated using the forward projection of the measured $\tau$ as a function of the compressor grating separation.

For pulse lengths greater than the transform limit, the pulses acquire a frequency chirp which for Gaussian pulses is defined as $a=\frac{1}{2}\psi''\tau^2$ where, $\psi(t)=\omega_0 t +\frac{at^2}{\tau^2}$, and $\omega_0$ is the central angular frequency of the IR pulse.

\subsection{Data acquisition and error analysis}
The IR laser pulses were measured with an Ophir power meter with 2\% rms uncertainty.
The employed THz detectors had a measurement uncertainty of 3\% rms. The shot-to-shot energy fluctuations of the the IR laser was 0.5\% rms corresponding to 1\% rms shot-to-shot energy fluctuations of the measured THz energies.  The measurements were averaged over 100 shots with an oscilloscope.  The error bars on our figures reflect the square of the sums for these uncorrelated errors.

\end{methods}



\bibliographystyle{naturemag}


\begin{addendum}
 \item The authors thank M. Kellermeier, C. Bosch, and P. Altmann (DESY) for engineering support. We thank C. Bruni (PHIL) for useful discussions and K. Floettmann for careful revision of the manuscript.  The authors are indebted to initial simulations, preliminary experiments and many discussions that were carried out with K. Ravi,  M. Hemmer, H. Olgun, N. Matlis and F. X. K\"{a}rtner within the AXSIS collaboration. This project was funded by the Laserlab-Europe Grant Agreement No. 654148, European Union’s Horizon 2020 Research and Innovation programme under Grant Agreement No. 730871 and also supported by the European Research Council (ERC) under the European Union’s Seventh Framework Programme (FP/2007-2013)/ERC AXSIS Grant agreement No. 609920.
 \item[Author contributions] F.L. developed the idea, lead the experiment and wrote the manuscript.  F.L., T.V., E.B., J.D., B.L., A.P., and M.P., carried out the experiment. Pulse duration measurements were performed by A.P. and M.P.  F.M. helped with data analysis and software development for the experiment. T.V., U.D., and R.A. contributed in management and procurement.  All authors contributed to improving the manuscript.
 \item[Competing Interests] The authors declare that they have no
competing financial interests.
 \item[Correspondence] Correspondence and requests for materials
should be addressed to Fran\c{c}ois Lemery (email:francois.lemery@desy.de).
\end{addendum}

\begin{table}
\centering
\caption{LN wafer specifications}
\medskip
\begin{tabular}{ccc}
\hline
parameter & value & unit \\
\hline
cut & +X $\pm$ 0.2 & deg \\
thickness & $\frac{\Lambda}{2}=300\pm 30$ & $\mu$m \\
AR-coating 700-900 nm & R$<$0.25 & \% \\
total thickness variation (TTV) & $<$10 & $\mu$m \\
percentage local thickness variation & $>$95 & \% \\
\hline
\end{tabular}
  \label{tab:wafers}
\end{table}

\begin{table}
\centering
\caption{Laser parameters of LASERIX}
\medskip
\begin{tabular}{ccc}
\hline
parameter & optimal values & used values \\
\hline
pulse energy & 1.2~J & 1.2~J \\
rms pulse duration & 50~fs & 50~fs to 20~ps \\
FWHM beam diameter & 45~mm  & 18 and 15~mm \\
super Gaussian order & 5 & 4 \\
repetition rate & 10~Hz & 10~Hz \\
\hline
\end{tabular}
  \label{tab:laserix}
\end{table}

\end{document}